\documentclass[
aps,
prl,
reprint,
superscriptaddress,
amsmath,
amssymb,
]{revtex4-1}

\usepackage{color}
\usepackage{amsmath}
\usepackage{amssymb}
\usepackage{amsfonts}
\usepackage{amsthm}
\usepackage{float}
\usepackage{bm}
\usepackage{latexsym}
\usepackage{hyperref}
\usepackage{multirow}
\usepackage{graphicx}
\hypersetup{
    colorlinks,
    citecolor=blue,    filecolor=blue,
    linkcolor=blue,     urlcolor=blue
}

\begin{document}

\title{Transverse instability in the "light sail" ion acceleration} %Title of paper

\author{Y. Wan}
\email[]{yang.wan@weizmann.ac.il}
\affiliation{Department of Physics of Complex Systems, Weizmann Institute of Science, Rehovot 7610001, Israel}

\author{I. A. Andriyash}
\affiliation{Department of Physics of Complex Systems, Weizmann Institute of Science, Rehovot 7610001, Israel}

\author{W. Lu}
%\email[]{weilu@tsinghua.edu.cn}
\affiliation{Department of Engineering Physics, Tsinghua University, Beijing 100084, China}

\author{W. B. Mori}
\affiliation{University of California Los Angeles, Los Angeles, CA 90095, USA}

\author{V. Malka}
\affiliation{Department of Physics of Complex Systems, Weizmann Institute of Science, Rehovot 7610001, Israel}
\affiliation{Laboratoire d'Optique Appliqu\'{e}e, Ecole polytechnique - ENSTA - CNRS - Institut Polytechnique de Paris, 828 Boulevard des Mar\'{e}chaux, 91762 Palaiseau Cedex, France}

%\thanks{}


\date{\today}

\begin{abstract}
Acceleration of ultrathin plasma foils by laser radiation pressure promises compact alternatives to the conventional ion accelerators. It was shown, that a major showstopper for such schemes is a strong transverse instability, which develops the surface ripples, and is often attributed to the Rayleigh-Taylor (RT) type. However, simulations indicate, that these perturbations develop the features, that cannot be consistently explained by the RT mechanism. Here we develop a three-dimensional (3D) theory of this instability, which shows that its linear stage is mainly driven by strong electron-ion coupling, while the RT contribution is actually weak. Our model provides the instability spectral structure and its growth rate, that agrees with the large scale 3D particle-in-cell simulations. Numerical modeling shows, that target destruction results from a rapid plasma heating induced by the instability field. Possible paths to instability mitigation are discussed.
\end{abstract}


\maketitle %\maketitle must follow title, authors, abstract and \pacs

\section{Introduction}
\begin{figure}[htbp]
\centering
\includegraphics[width=0.8\linewidth]{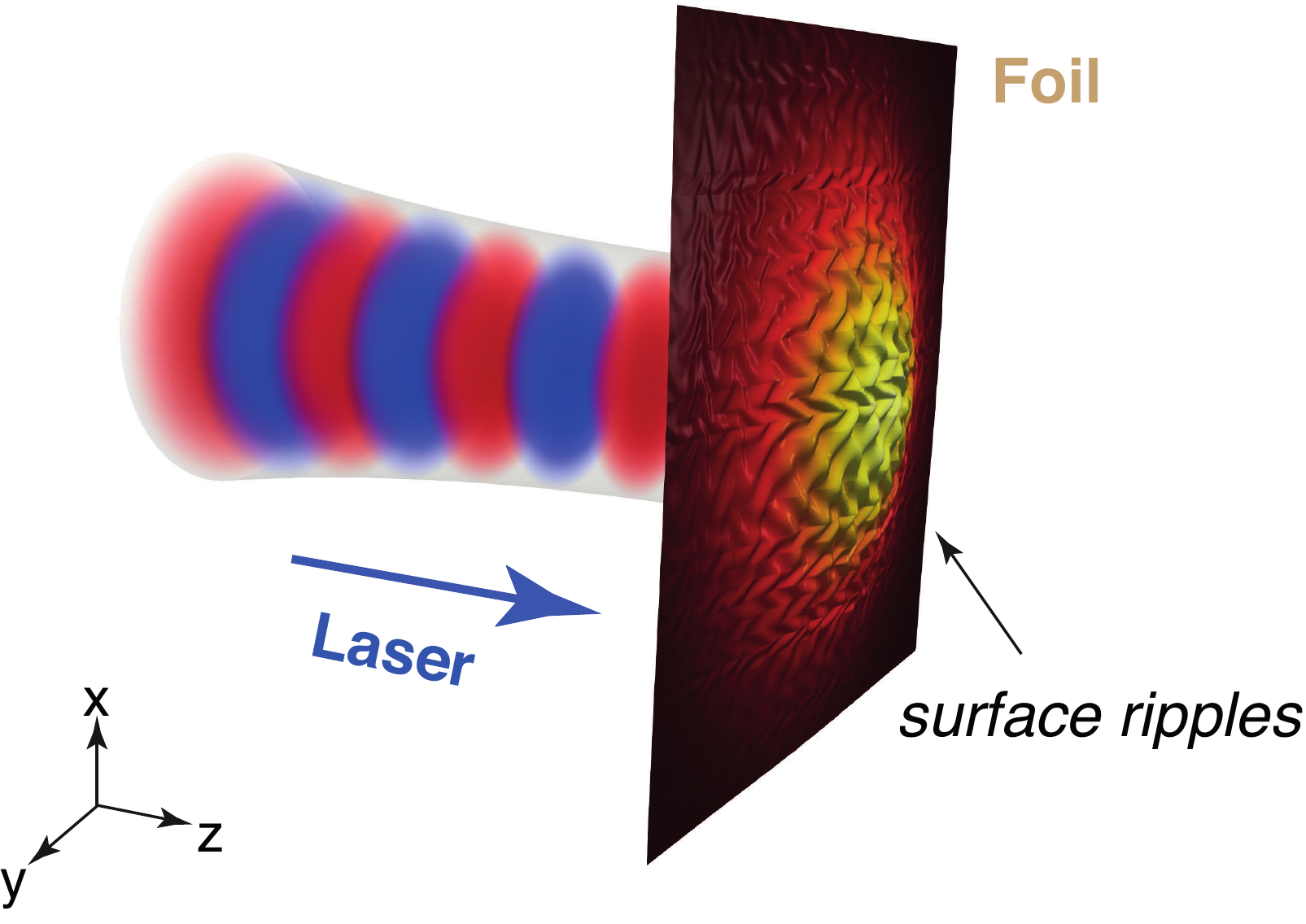}
\caption{\label{schematic_drawing} The schematic drawing of transverse instability induced in the light sail regime.}
\end{figure}

Interacting with a thin foil, an ultra-intense laser pulse can accelerate ions on a very short distance, which makes such schemes attractive for the applications ranging from radiography of high-energy density targets \cite{borghesi2002PIprl, mackinnon2006pi, li2006pi} to radiotherapy \cite{bulanov2002CTpra, malka2004ct}. Among various acceleration mechanisms, one promising option is to push a solid target directly by the radiation pressure force \cite{esirkepov2004, macchi2005, zhang2007a, robinson2008, klimo2008, yan2008, macchi2009}. In the case of a very thin nanofoil, it can be pushed as a single sheet, which is known as the ``light sail'' (LS) acceleration \cite{macchi2009}. Ideally, in a one-dimensional LS process, efficient and quasi-monoenergetic ion acceleration is predicted by using a normally incident circularly polarized laser, where $\mathbf{J} \times \mathbf{B}$ heating is strongly suppressed \cite{macchi2005}. In the realistic conditions, the surface deformations from a nonuniform laser profile enable more efficient plasma heating \cite{brunel1987}. Tailoring the laser and target profiles allows to create a uniform radiation pressure distribution, preventing the macroscopic surface bending \cite{chen2008, chen2009prl}. However, the local surface distortions produced by the plasma instability, turn out to be a greater problem (see Fig.~\ref{schematic_drawing}).

Simulations and experiments indicate a strong transverse instability in LS process, which generates surface and density perturbations of the plasma foil \cite{klimo2008, robinson2008, chen2009prl, palmer2012}. This process is not yet well understood, and the mechanism of Rayleigh-Taylor (RT) instability driven by the light pressure was suggested as a possible  explanation \cite{ott1972nonlinear,pegoraro2007}. Simulations indicate, that the developing fluctuations have a well pronounced periodic structure, while the property of RT is to grow faster for the shorter wavelengths (in absence of stabilizing mechanisms). In \cite{eliasson2015,sgattoni2015}, this contradiction was addressed by suggesting, that the scale of the produced surface ripples is imposed by the diffraction patterns of the laser field on the surface, and therefore should be close to the laser wavelength. In its turn, this consideration also contradicts the numerical simulations, where the modulation scales were shown to vary with different laser plasma parameters \cite{zhang2011,wan2016}. More recently an alternative explanation suggested, that this instability does not origin from RT process, but is induced by a strong coupling between oscillating electrons and quasi-static ions \cite{wan2016,wan2018b}.

In this paper we propose a generalized model that includes all possible effects in the full three-dimensional (3D) geometry. The new theory allows to self-consistently account for the RT development, and compare the contributions of both mechanisms. For the first time we obtain analytically the instability growth rate and its mode structure, which are fully consistent with the 3D particle-in-cell (PIC) simulations. In the second part of the paper we show, that the instability itself does not destroy the target, but saturates when entering nonlinear stage. During this stage, the density fluctuations induce a rapid plasma heating, which triggers the merging of the developed small structures and finally breaks the whole target thus letting the laser penetrate through the plasma. In the final part of the study, we discuss the possibilities to minimize the instability effect following from the model.

\section{Three-dimensional instability mechanism}

\subsection{Numerical demonstration}

Let us start the discussion by showing a typical LS scenario in a highly resolved 3D PIC simulation, using a fully relativistic electromagnetic code OSIRIS \cite{fonseca2002}. In this simulation, a circularly polarized (CP) laser pulse with a frequency $\omega_0$ and normalized amplitude $a_0=5$ is launched at $t=0$ along $z$-axis, from the left boundary of the simulation domain with the grid resolutions of $0.04\,c/\omega_0$, $0.04\,c/\omega_0$ and $0.02\,c/\omega_0$ in $x$, $y$ and $z$ directions, where $x$, $y$ represent transverse directions and $z$ is the longitudinal axis; $c$ is the speed of light in vacuum. For a more accurate comparison with the following theoretical analysis, we consider a laser with the flat-top temporal profile bounded by $15\,c/\omega_0$ rise and fall ramps, and uniform transverse profile. The foil target is placed at $z=5\,c/\omega_0$, and is modeled as a pre-ionized uniform electron-proton plasma with the thickness $l_0=c/\omega_0$, density of $10n_c$. Plasma is initialized with equidistantly distributed macro-particles -- 64 particles per species per cell, and with electron temperature of 10 eV, which provides initial perturbations to seed the instability.

\begin{figure}[htbp]
\centering
\includegraphics[width=\linewidth]{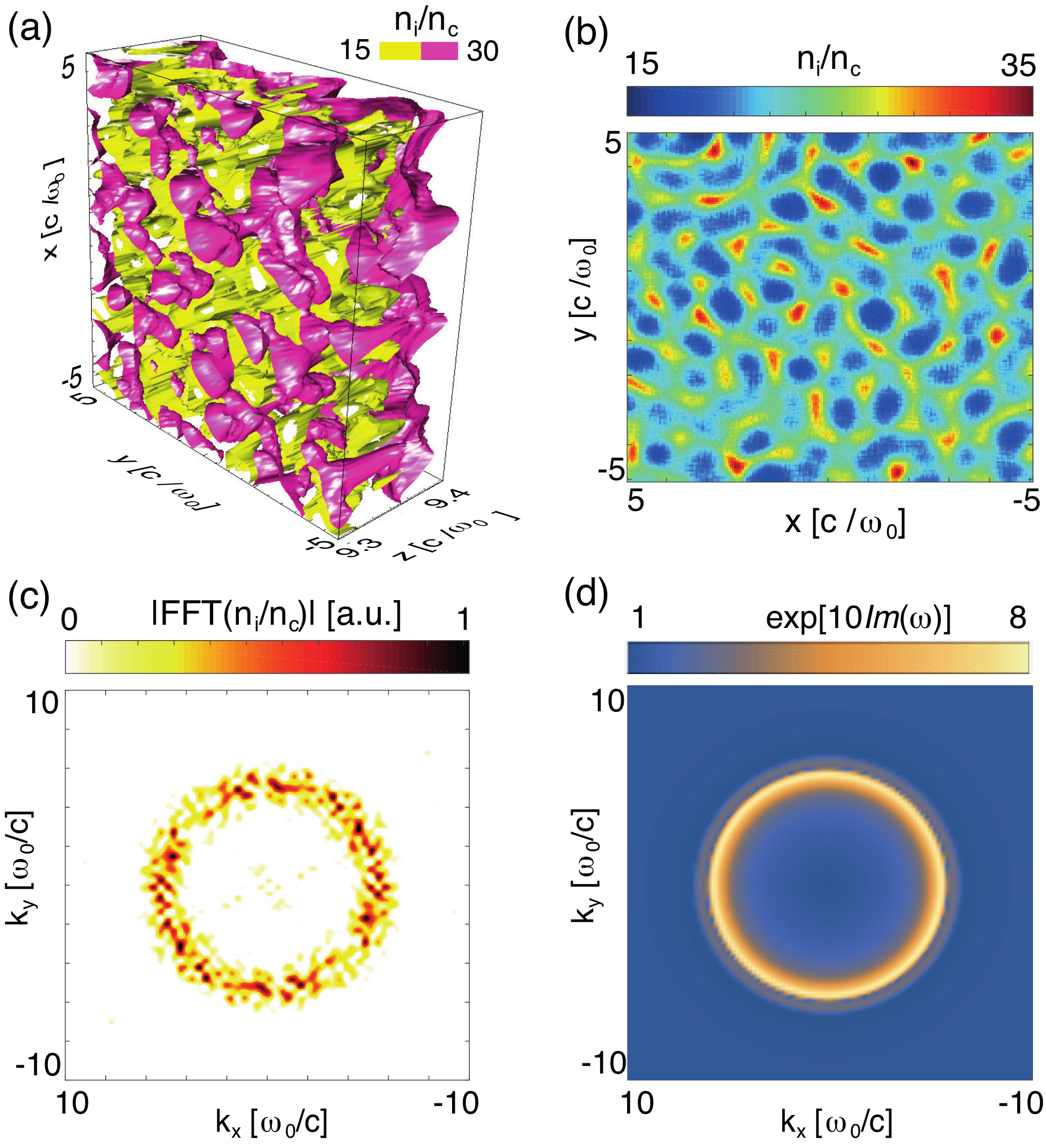}
\caption{\label{3d_ion_density} (a) The 3D distribution of the proton density within the region of FWHM at $t=60~\omega_0^{-1}$.
(b) The averaged ($x-y$) slice distribution of proton density in (a).
(c) The FFT of the proton density shown in (b).
(d) The corresponding ($k_x$, $k_y$) space calculated numerically from dispersion relation.}
\end{figure}

\begin{figure*}[htbp]
\centering
\includegraphics[width=0.8\linewidth]{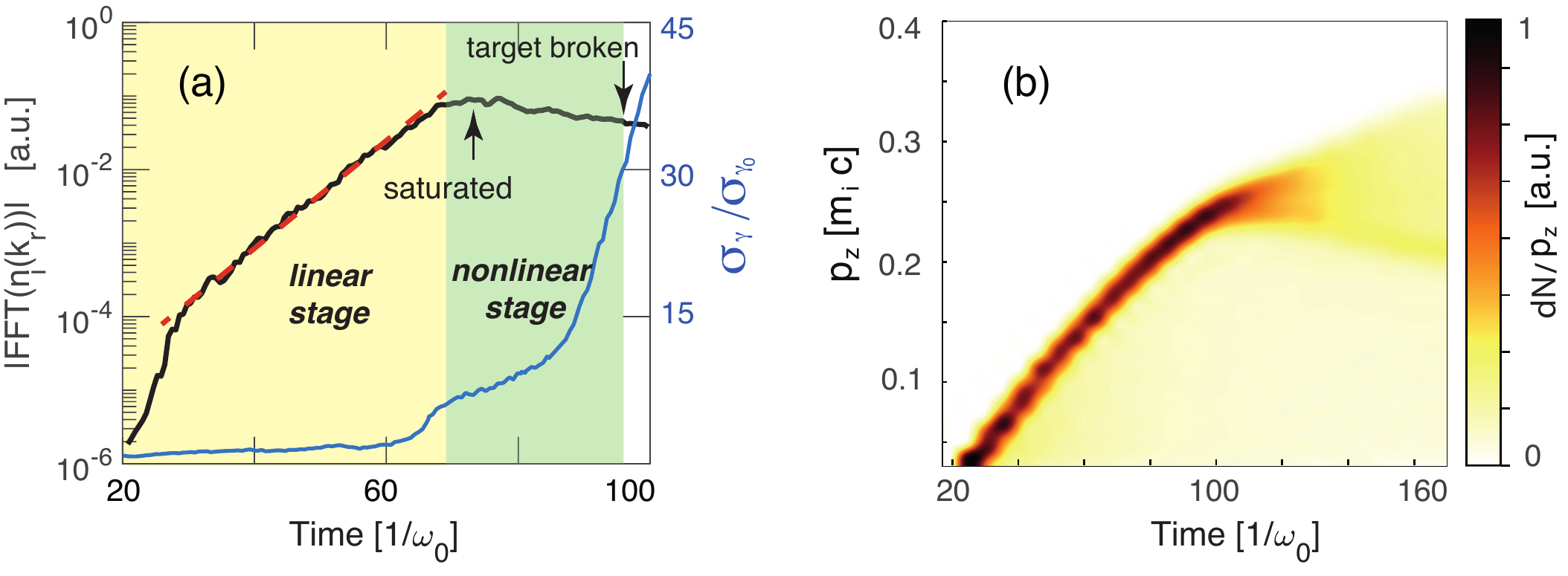}
\caption{\label{time_evolution}  (a) Spectral amplitude of the proton density fluctuations (black) and electron energy spread (blue) as the functions of interaction time. The red dashed line represents the fitting curve of exponential growth.
(b) Temporal evolution of the proton longitudinal momentum density. Here $\sigma_{\gamma0}$ defines the initial electron RMS energy spread.}
\end{figure*}

Fig.~\ref{3d_ion_density}a shows a 3D distribution of proton density within the region of its full width half maximum (FWHM), at $t=60~\omega_0^{-1}$, when the foil is pushed by the laser pulse. At that time one can see, that both target density and shape are strongly modulated, and the average proton density in the compressed foil reaches, ${n_p \simeq 23 n_c}$. The ($x-y$) density distribution averaged along $z$ is plotted in Fig.~\ref{3d_ion_density}b, demonstrating a quasi-regular structure of the emerging perturbations. This structure can also be analyzed in the spectral space, and in Fig.~\ref{3d_ion_density}c we show the 2D Fourier transformation of Fig.~\ref{3d_ion_density}b. The unstable mode has a ring structure with the central radial wavenumber $k_m = 5.3\,\omega_0/c$, and its amplitude can be used to describe instability dynamics, as shown by a black curve in Fig.~\ref{time_evolution}a. At early times, $t\lesssim 25 c/\omega_0$, initial thermal energy of electron plasma provides a quick rise of charge density fluctuations \cite{akhiezer1962contribution}. After that, the resonant mode $k_m$ is excited and grows exponentially, $\delta n\propto \exp(\gamma_m t)$, with the rate $\gamma_m = 0.18\,\omega_0$ (red dashed line in Fig.~\ref{time_evolution}a).

The electron energy spread shown with the blue curve in Fig.~\ref{time_evolution}a, characterizes plasma heating dynamics. One can see, that during the exponential instability growth, the heating is effectively prevented. Instability gets saturated at $t\gtrsim 70 c/\omega_0$, and from this moment the electron energy spread starts to grow. Rapid heating disperses electrons around the target, and leads to the target breaking at $t\simeq 100 c/\omega_0$. This final stage is also observed in proton acceleration dynamics shown in Fig.~\ref{time_evolution}b. During the linear and saturation stages, acceleration is uniform, and results in a quasi-monoenergetic proton spectrum. At the final stage, the foil surface is no longer sharp, and does not support the radiation pressure acceleration. At that point, the acceleration is governed by the plasma expansion process, which drastically increases the proton spectral bandwidth, washing off the quasi-monoenergetic feature.

\subsection{Analytic model}

Let us now describe the linear stage of instability using the relativistic two-fluid equations of electrons and ions. For the derivation simplicity we consider a reference frame, which follows the accelerated plasma foil. This frame is non-inertial as it experiences acceleration, ${\alpha_\text{in}\equiv \mathrm{d}v_{in}/\mathrm{d}t = P_{rad}/(m_i n_0 l_0)}$, where $P_\text{rad} = 2 I_\text{las}/c$ is the radiation pressure produced by a laser with intensity $I_\text{las}$, $m_i$ is the ion rest mass, $n_0$ and $l_0$ correspond to the non-perturbed target density and thickness respectively. In this analysis we consider only electrostatic fluctuations along the laser polarizations, and neglect the small contributions of electromagnetic modes \cite{wan2018b}.

The continuity and transverse motion equations for the cold electron and ion fluids read:
\begin{subequations}\label{fulldescrip}
\begin{eqnarray}
 &\cfrac{\partial n_s}{\partial t} + \nabla \cdot (n_s \mathbf{v}_s)=0 \,, \label{flu_con}\\
 &\cfrac{\partial \mathbf{p}_{\perp\,,s}}{\partial t} + \mathbf{v}_s\cdot\nabla \mathbf{p}_{\perp\,,s} = q_s \mathbf{E}_\perp \,, \label{flu_mon_trans}
\end{eqnarray}
where $\mathbf{v}_s$, $\mathbf{p}_s$, $n_s$ and $q_s$ denote respectively the velocity, momentum, density and charge of the specie $s$ (either electrons or ions). In the accelerating frame, the longitudinal ion motion is governed by the pressure $P_\text{rad}$ and the inertia force $m_i\alpha_\text{in}$,
\begin{equation}
 \cfrac{\partial p_{z\,,i}}{\partial t}=\cfrac{P_\text{rad}}{n_i l_0}-m_i\alpha_\text{in}\label{flu_mon_lon}\,,
\end{equation}
and the charge densities of the ion and electron fluids are coupled via the Poisson equation:
\begin{equation}
  \nabla\cdot \mathbf{E}=4\pi (q_in_{i}-en_{e})\label{flu_pos}\,.
\end{equation}
\end{subequations}

To simplify further derivations, we make a few assumptions. Firstly, let us consider ions to be non-relativistic, which is well justified at the beginning of the LS process, when the instability is already developing, while particle energies are still relatively low. We also note, that surface distortions produced by the instability have typically much greater size than the thickness of the compressed foil. This allows us to assume plasma to be infinitely thin, and consider all related quantities to only depend on ($x$, $y$, t). Following the linear perturbation theory, we divide all field and plasma quantities into the zero- and first-order parts, ${f=f_0+\tilde{f}}$, and neglect the terms $\propto\tilde{f}^2$ and $\tilde{f}^3$. Electric fields of a CP laser ${E_{x0}=E_0\sin(\omega_0 t)}$ and ${E_{y0}=E_0\cos(\omega_0 t)}$, and the corresponding electron oscillations define the zero-order values of the transverse field and electron velocity $v_{e0} = v_{\perp\,e\,0}$ respectively, where $\omega_0$ is the laser frequency in the co-moving frame.

We also note, that the transverse force in Eq.~(\ref{flu_mon_trans}), corresponds to a case of the ideally flat foil. Adding the surface distortions $\tilde{z}$, produces the transverse force, ${-P_\text{rad}\nabla_\perp \tilde{z}}$, in the first order component of Eq.~(\ref{flu_mon_trans}) for electrons:
\begin{equation}\label{flu_mon_trans_mod}
 \cfrac{\partial \tilde{\mathbf{p}}_{\perp\,,e}}{\partial t} + e \tilde{\mathbf{E}}_\perp  = - \mathbf{v}_{e0}\cdot\nabla \tilde{\mathbf{p}}_{\perp\,,e}  - P_\text{rad}\nabla_\perp \tilde{z} \,.
\end{equation}
The two terms on the right hand side of Eq.~(\ref{flu_mon_trans_mod}) represent the contributions of the electron-ion coupled parametric instability \cite{wan2016}, and the Rayleigh-Taylor (RT) instability driven by the the radiation pressure coupled to the foil distortions \cite{ott1972nonlinear,pegoraro2007} respectively.

After all mentioned considerations, a linear set of equations can be obtained for the first-order terms for both fluids. Eliminating the terms $\tilde{n}_i$, $\tilde{\mathbf{v}}_i$, $\tilde{\mathbf{p}}_i$, $\tilde{\mathbf{E}}_{\perp}$ and $\tilde{z}$, the equations for electron density and momenta fluctuations can be obtained. Assuming further, that all quantities are ${\propto \exp[i(k_x x + k_y y - \omega t)]}$, we can linearize the derivatives, and finally obtain the equations for the Fourier components of $\tilde{n}_{e}$, $\tilde{p}_{x\,,e}$ and $\tilde{p}_{y\,,e}$ (in following we suppress the subscript $e$):
\begin{eqnarray}\label{disp_eqs}
&  \Omega_g \tilde{n}_{\omega+\omega_0} + \Omega_g^* \tilde{n}_{\omega-\omega_0} +  n_0 \kappa  \left( k_x \tilde{p}_{x,\,\omega}+k_y \tilde{p}_{y\,,\omega}\right) = \omega \tilde{n}_{\omega} \nonumber \,,\\
& \Omega_g\tilde{p}_{x,\,\omega+\omega_0} + \Omega_g^*\tilde{p}_{x,\,\omega-\omega_0} + k_x \tilde{n}_{\omega} \chi(\omega) =\omega \tilde{p}_{x,\,\omega}\,,\\
& \Omega_g\;\tilde{p}_{y,\,\omega+\omega_0}  + \Omega_g^* \tilde{p}_{y,\,\omega-\omega_0} + k_y \tilde{n}_{\omega} \chi(\omega)= \omega \tilde{p}_{y,\,\omega}\,,\nonumber
\end{eqnarray}
where we denote $\Omega_g=v_{os}(k_x-ik_y)/2$, ${\kappa=(2-v_{os}^2)/2\gamma_0}$, with $v_{os}=P_{os}/\gamma_0$ being the electron quiver velocity amplitude and $\gamma_0$ is the electron's zero-order Lorentz factor. We have also defined here a dispersion function:
\begin{equation}\label{disp_func}
\chi(\omega) = \cfrac{\alpha_{in}^2 m_i \omega_{pi}^2}{n_{e0}\omega^2 (\omega_{pi}^2-\omega^2)} \;-\; \cfrac{4\pi\omega^2}{(k_x^2+k_y^2)(\omega_{pi}^2-\omega^2)} \,,
\end{equation}
where $\omega_{pi}$ is the ion plasma frequency. The first term in Eq.~(\ref{disp_func}) is responsible for the RT process, and the second one defines the electron-ion coupled instability. In the expressions above, mass is in units of electron rest mass $m_e$, velocity in units of $c$ and momentum in units of $m_e c$.

Equations~(\ref{disp_eqs}) are written for a mode $\omega$, and they couple it with the up- and downshifted modes $\omega\pm\omega_0$. Following the standard parametric analysis \cite{kruer2018physics}, we write these equations for the modes ${\omega\pm\omega_0}$, and, neglecting the off-resonant terms ${\omega\pm 2\omega_0}$, obtain a closed set of equations for $\tilde{n}$, $\tilde{p}_{x}$ and $\tilde{p}_{y}$ for the modes $\omega$, ${\omega+\omega_0}$ and ${\omega-\omega_0}$. Since these equations are homogeneous (have no constant terms), the non-trivial solution exists only for the coefficient matrix with a zero determinant, which gives a dispersion relation between $\omega$ and $k$. More details on the derivation can be found in the Supplementary Material \cite{SuppMat}.

The resulting dispersion equation can be solved numerically, and in Fig.~\ref{3d_ion_density}d we show distribution of the imaginary part of $\omega$ (i.e. instability growth rate) in ($k_x$, $k_y$) space. For the coefficients in Eqs.~(\ref{disp_eqs}) we use the values of the electron and ion densities $\bar n_{e}=55~n_c$ and $\bar n_i=23~n_c$ extracted from the simulation and averaged in the FWHM region, and $\bar \gamma_0=2.2$ is calculated from the simulated electron spectrum. Figures ~\ref{3d_ion_density}c and \ref{3d_ion_density}d clearly demonstrate the similar ring structures, and the simulated radius value $5.3\,\omega_0/c$ has a good agreement with the theoretical value $5.8\,\omega_0/c$. The growth rate given by the maximal imaginary part of this numerical solution is $\gamma_m = 0.21\,\omega_0$, also in good agreement with the simulated value (see Fig.~\ref{time_evolution}a). Considering separately two terms in Eq.~(\ref{disp_func}), one can estimate the contributions of each effect in the instability processes: $\gamma_{ei}= 0.2\,\omega_0$ is a growth rate of electron-ion coupling instability and $\gamma_{RT}=0.06\,\omega_0$ corresponds to the RT process.

Making a few more assumptions for the numerical solution of the dispersion equation (similar to ones in section III of \cite{wan2018b}), we obtain simpler expressions for the maximal growth rate:
\begin{subequations}
\begin{equation}\label{gamma_exp}
 \gamma_m\approx\sqrt{\gamma_{ei}^2+\gamma_{RT}^2}\,,
\end{equation}
and the wavenumber of the fastest growing mode:
\begin{equation}\label{km_exp}
k_{xm}^2+k_{ym}^2\simeq2\kappa\frac{\omega_{pe}^2}{v_{os}^2}\,,
\end{equation}
\end{subequations}
where the contributions of two instabilities to the growth rate are:
\begin{eqnarray}    
&& \gamma_{ei}\simeq 2 \left(\omega_{pi}^2\omega_{pe}\right)^{1/3} \left(\kappa m_e / m_i \right)^{1/6}\,,\nonumber \\
&& \gamma_{RT}\simeq \left( \sqrt{\kappa/2}\; \alpha_{in} \omega_{pe} / v_{os} \right)^{1/2}\,,\nonumber
\end{eqnarray}
Equation~(\ref{km_exp}) defines an isotropic mode structure in the ($k_x$, $k_y$) space in the form of a ring with the radius of ${k_{r}=\sqrt{2\kappa}\omega_{pe}/v_{os}}$. In the 1D case, $k_{y}=0$, Eq.~(\ref{km_exp}) is reduced to the expression for $k_m$ obtained in Ref \cite{wan2016}. In a case $\gamma_0\gg1$, Eq.~(\ref{km_exp}) can be also simplified as ${k_{r}\simeq\omega_{pe}/c\sqrt{\gamma_0}}$, corresponding to the relativistic plasma skin depth. In most cases, $\gamma_{ei}$ is much larger than $\gamma_{RT}$, indicating the former dominates the whole instability development. For a low laser intensity ($a_0<1$), $n_e$ is close to the value of $n_i$, and the expression of $\gamma_{ei}$ is simplified as $2\omega_{pi}$.

Note, that for non-relativistic interaction, the expressions Eqs.~(\ref{km_exp}) and (\ref{gamma_exp}) can be directly applied in the laboratory reference frame, while for the higher ion energies these estimates should be Lorentz-transformed.

\section{Nonlinear instability evolution}

In Fig.~\ref{time_evolution}, we see that after the linear instability phase its exponential growth saturates. Meanwhile, electron heating starts, and until certain level it does not affect acceleration of ions and their spectral bandwidth. As electron temperature grows, particles spread further from the target, and density of electron plasma decreases. At that point laser penetrates through the target, and the LS process can no longer continue. This moment we call the \textit{target breaking}, and for the optimal experiment design it is important to understand this process better.

\begin{figure}[htbp]
\centering
\includegraphics[width=0.95\linewidth]{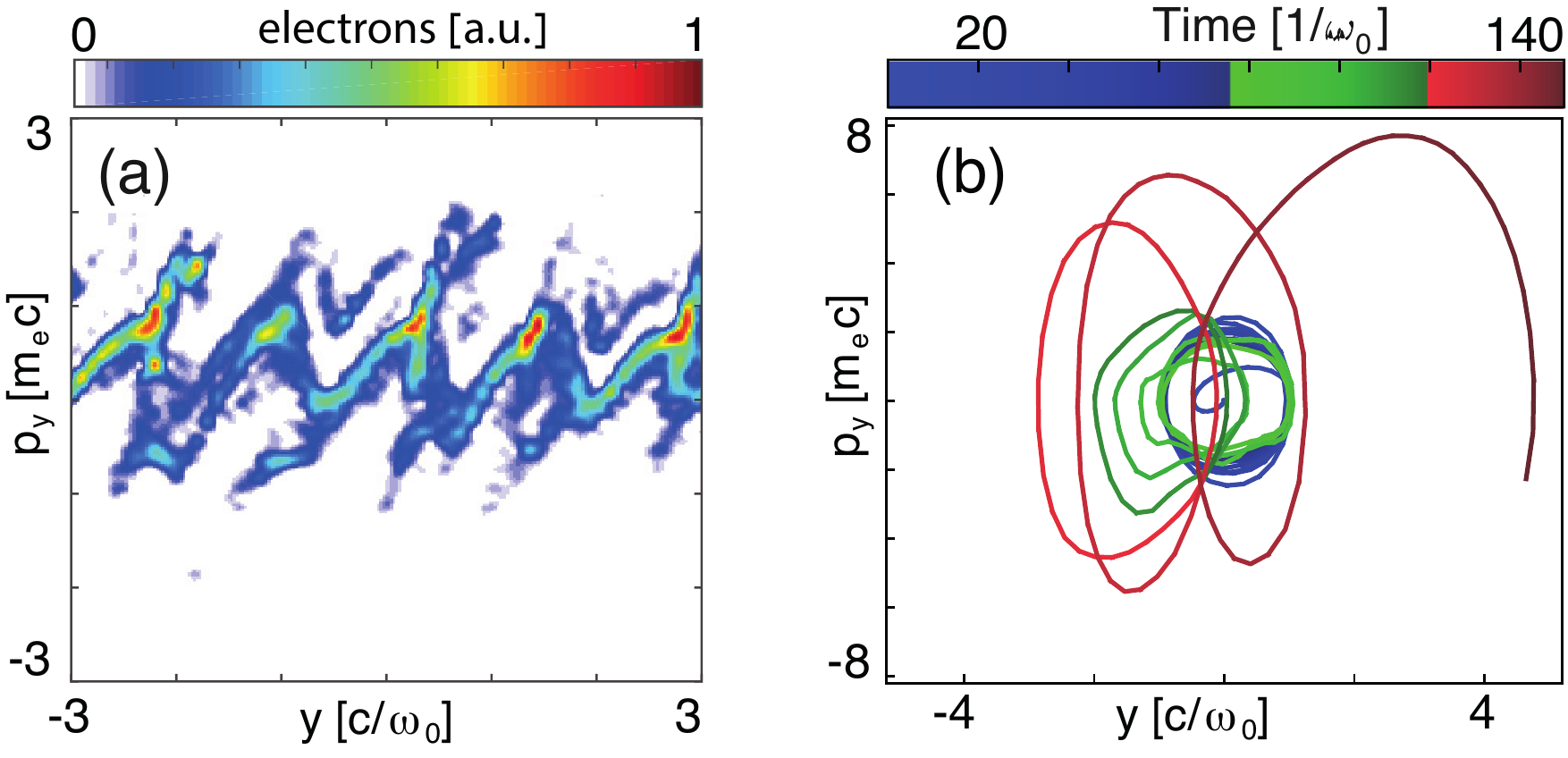}
\caption{\label{electron_heating} (a) The modeled electron transverse phase space ($y-p_y$) after the instability saturation. (b) Transverse phase orbit of a sampled electron from the same 2D PIC simulation.}
\end{figure}

To explore the induced electron heating, we perform a 2D PIC simulation with the same physical parameters as before, but finer resolutions $\Delta y=\Delta z=0.02\,c/\omega_0$ and 80 particles per species in each cell. In Fig.~\ref{electron_heating}a we look at the electron transverse phase space ($y-p_y$) at the time when instability saturates. Here we see the periodic transverse modulations produced by the instability, and at the peaks of this mode one can see the particles escaping the modulated structure, which corresponds to the wave-breaking process. These higher energy electrons travel between the modulation nodes, and their dynamics can be observed by following a sample electron trajectory, as shown in Fig.~\ref{electron_heating}b. During the linear instability phase (blue part), electron oscillates around its initial position in the laser field with a transverse momentum $p_y\simeq 2 m_e c$, defined by the partially screened laser field. At the saturation stage (green part), the instability field becomes strong enough to force electron to drift away from its initial position along the surface. Finally, near the target breaking (red part) laser field fully penetrates through the plasma, and electron can reach the momentum up to $6~m_e c$.

As we see, the instability saturation time is related to the modulation wave-breaking. At the linear stage, the  first-order perturbations of electron motion are confined in the local oscillating electric field with a typical wave number $k_m$. These perturbations can be expressed as $\tilde{p} \simeq \Delta p\sin{\omega_0t}\sin{k_m y}$, where $\Delta p$ is the amplitude of the first-order momentum. As instability develops, $\Delta p$ grows exponentially, and so does the coordinate oscillation amplitude $\Delta y\sim \Delta p/m_e\kappa\omega_0$, until it reaches the plasma potential width, $k_m\Delta y\sim1$. At that point, electrons start to escape the unstable plasma wave, which corresponds to the wave-breaking process. Combining this relation with Eqs.~(\ref{disp_eqs}), one can estimate the saturated density perturbations $\Delta n_\text{sat}$, as
\begin{eqnarray}\label{sat_amp}
\frac{\Delta n_\text{sat}}{n_0}\simeq \frac{\Delta p_\text{sat}}{p_0}\simeq \frac{2\,\omega_{0}\,\gamma_0^{\frac{7}{2}}}{\omega_{pe}\,(\gamma_0^2+1)^{\frac{3}{2}}}\,.
\end{eqnarray}
Eq.~(\ref{sat_amp}) predicts that plasma heating onsets, when $\Delta n_e/n_{e}\simeq 0.3$, which compares reasonably well with the simulation value of 0.2 at saturation. We note, that at this time, density fluctuations are still small and the target is opaque to the laser, such that the LS process is maintained well.

\begin{figure}[htbp]
\centering
\includegraphics[width=0.95\linewidth]{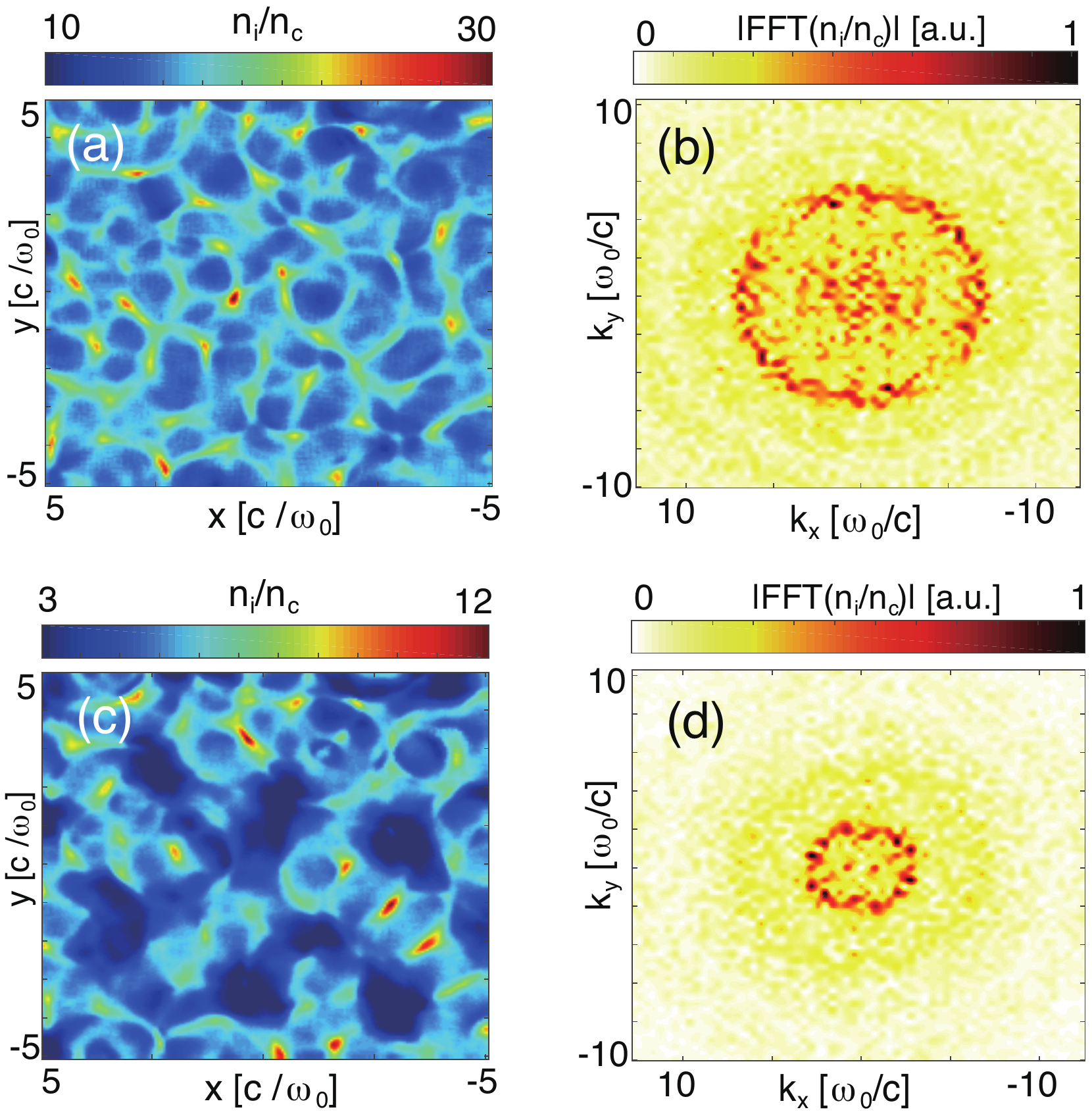}% Here is how to import EPS art
\caption{\label{nonlinear_proton} The proton density distribution (a,c) and the corresponding Fourier transforms (b,d) at two different times $t_1=75~\omega_0^{-1}$ (a,b) and $t_2=100~\omega_0^{-1}$ (c,d).}
\end{figure}

Let us now consider more carefully the development of the saturation stage. When an electron obtains enough energy to escape the unstable plasma wave, it travels along the target surface. The fields of the nodes may slightly differ from each other (see Figs.~\ref{3d_ion_density}b and \ref{nonlinear_proton}a), hence, escaping one node electron may still be trapped by another one with a higher field. This stochastic process produces electron flow from smaller to bigger wave buckets, and lead to the nodes merging. In the spectral space, this process corresponds to a shift of the developed unstable plasma mode to the longer wavelengths. This spectral down-shift driven by thermalization is known in physics of turbulent plasma instabilities \cite{tsytovich2012nonlinear}, and it can also be observed in our numerical simulations. In Fig.~\ref{nonlinear_proton}, we show the proton density maps and their Fourier images at two different times of $75~\omega_0^{-1}$ and $100~\omega_0^{-1}$. One may clearly see, that during the saturation phase, the sizes and the spacings between the density perturbations increase. In the spectral space the long-wave fluctuations develop within the initial ring structure, with the unstable $k$ value falling from 5 $\omega_0/c$ to about 1.5 $\omega_0/c$. In this simulation, the maximum acceleration time $t_\text{acc}$ is around $100~\omega_0^{-1}$, corresponding to about 40 fs for a 800nm laser.

Besides the typical simulation shown above, we have also systematically scanned over a wide range of laser plasma parameters to verify the validity of the theoretical model. Moreover, since the derivation makes assumption of a homogeneous radiation pressure, it is also important to see how instability development is affected by a more realistic laser profile. For this, we have also performed a 3D PIC simulation with a transverse Gaussian profile, that exhibits similar ring structures in the central region within the laser focal spot. For the case of a tightly focused laser, we have observed additional electron heating arousing at the target periphery, which may spread and destroy the target even before the instability fully develops. More details of these additional simulations are presented in the Supplementary Material \cite{SuppMat}.

\section{Discussion}

As we have shown, the plasma light-sail instability triggers the target heating and indeed presents a major obstacle for efficient ion acceleration. It also follows, that this instability always develop in the laser-driven thin foils, and cannot be avoided except by preventing the coupling between the oscillating electrons and ions. One such possibility is to physically separate electrons and ions using, for example, a tweezer effect of two colliding lasers with the different frequencies \cite{wan2017stable}. In this case, however, acceleration is no longer provided by the radiation pressure in a normal manner. Considering only LS regime, we will look to minimize the instability growth rate, and for this it is crucial to identify the proper interaction parameters.

\begin{figure}[htbp]
\centering
\includegraphics[width=0.75\linewidth]{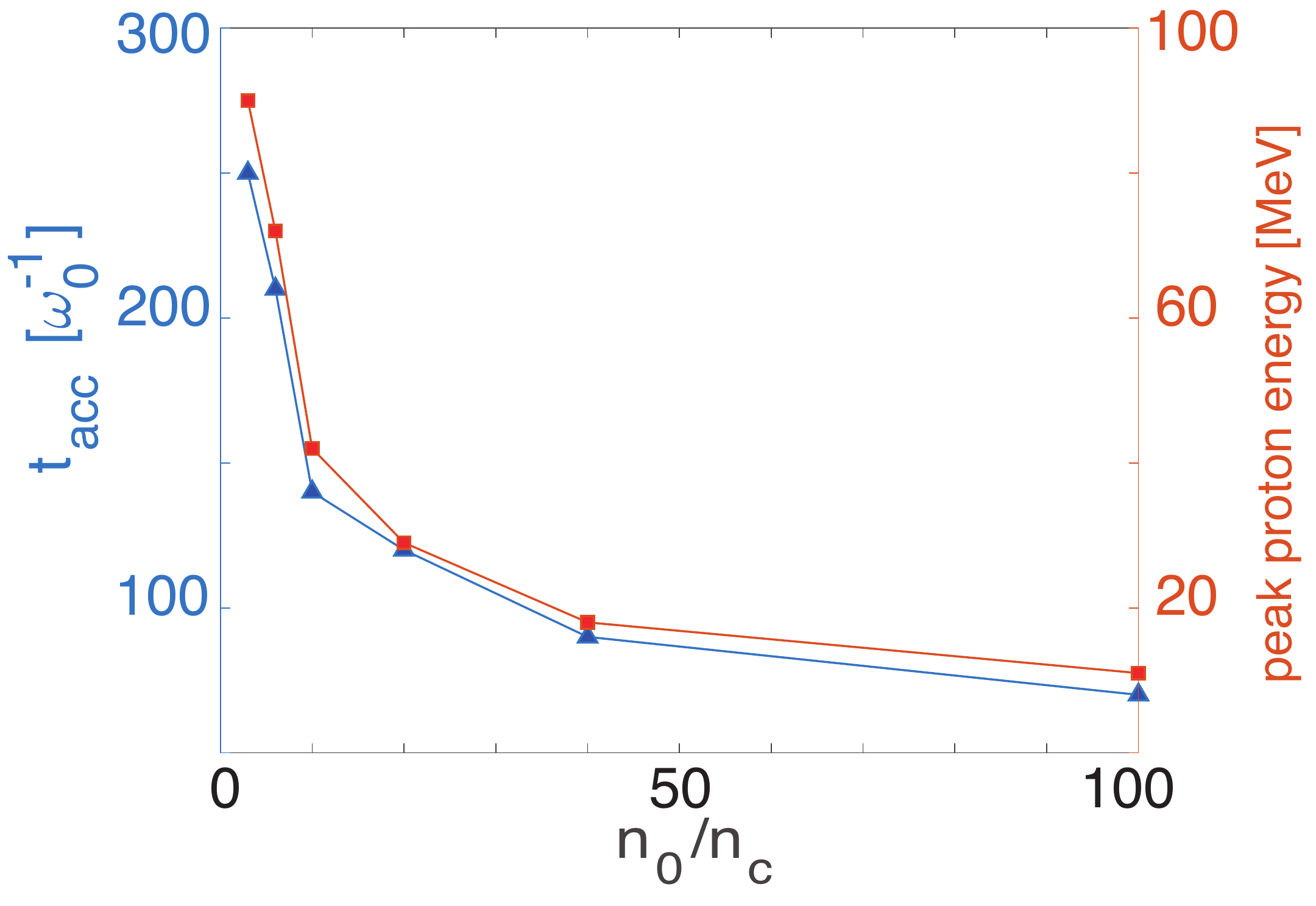}% Here is how to import EPS art
\caption{\label{varying_density} Acceleration time $t_\text{acc}$ and final proton energy as the functions of initial target density for a fixed $a_0=5$, and thickness adjusted  as $l_0 = 2a_0 n_c c /(n_0\omega_0) $.}
\end{figure}

Let us consider the acceleration efficiency as a function of the target design, i.e. the ion compound, initial density or temperature. As mentioned before, in the optimal LS process the foil thickness, density and laser amplitude are coupled by the balance condition ${2a_0\simeq l_0 \omega_0 n_0/c n_c}$ \cite{yan2008,macchi2009}. Therefore, for the given laser characteristics an optimum can be found by varying a single parameter -- either foil thickness or density. From Eq.~(\ref{gamma_exp}) follows, that instability growth correlates with the plasma density, which in its turn correlates with the initial target density. This means, that for the lower target density the acceleration time can be extended resulting in the higher proton energies. This statement can be verified numerically, and here we perform a series of 2D PIC simulations for $a_0=5$ and varying foil density, while its thickness is adjusted with respect to the balance condition. In Fig.~\ref{varying_density} we show the times before foil breaking and ion energy corresponding to the peak in their quasi-monoenergetic spectra. This result clearly confirms our hypothesis, as when the target density decreases from 100 $n_c$ to 3 $n_c$, maximum acceleration time is extended from $70~\omega_0^{-1}$ to around $250~\omega_0^{-1}$, and the final proton energy increases from 12 MeV approaching 90 MeV. We should note here, that in 2D case, instability has less degrees of freedom and develops slightly slower, while the behavior with respect to the parameters remains equivalent. %Practically, the available thin foils always provide the solid-density plasma, $n_p\gtrsim 100 n_c$, while the lower density, near-critical targets require a more complicated fabrication process. Presently such targets are possible using the foam materials, carbon nanotubes or cryogenic hydrogen jets, however, achieving sub-micrometer thickness for these targets may still be a challenge.

One more option to inhibit the instability, is to consider an initial electron temperature $T_e$ high enough to provide a thermal smoothing of the perturbations \cite{wu2014}. The smoothing scale is defined by the electron Debye length $\lambda_D=\sqrt{T_e/4\pi n_0e^2}$, and to affect the instability it needs to be of the order of the instability scale length, i.e. $k_m\lambda_D\sim 1$.  In the case of an intense laser, this condition defines temperature $T_e\sim \gamma_0$, which can be as high as several MeV and leads to a rapid plasma expansion. An additional numerical study of the electron temperature effect have demonstrated, that in all cases it leads to an earlier target breaking than in a case of the initially cold plasma.

Among other possibilities to improve the LS performance, it is worth to mention the multi-species targets, where heavy ions provide a co-moving electrostatic field for protons even after the plasma is disrupted by the instability \cite{yu2010stable, qiao2010radiation, kar2012ion, liu2013generation, wan2018tri}. Another proposed alternative is to use an extremely intense single-cycle laser to provide acceleration on a time before the instability develops \cite{zhou2016proton}.

In conclusion, we have presented and explored a three-dimensional theory of the transverse instability in the laser-driven plasma light sail. It is shown, that both strong electron-ion coupling effect and nonuniform laser pressure contribute to the instability at early linear stage, while the former dominates. We have identified the unstable mode structure and estimated its growth rate, which was found in good agreement with 3D PIC simulations in a wide range of parameters. The nonlinear instability stage was studied numerically, and it was shown that it saturates with the wave-breaking effect, and leads to a strong plasma heating. The following target breaking is accompanied by merging of the smaller density structures into the larger ones. Possible solutions to suppress the instability are discussed and summarized.

This work was supported by NSFC Grant No. 11425521, No. 11535006, No. 11475101 and No. 11775125, the Thousand Young Talents Program, Gerry Schwartz and Heather Reisman, Israel Science Foundation, and VATAT supports. Simulations were performed on Edison and Cori clusters at NERSC.

\bibliography{refs_for_TI_paper}
\end{document}